# Electrostriction Effects During Defibrillation

Michelle M. Fritz
Department of Physics
University of Michigan
Ann Arbor, Michigan 48109 USA

Phil W. Prior and Bradley J. Roth
Department of Physics
Oakland University
Rochester, Michigan 48309 USA

ABSTRACT

Background—The electric field applied to the heart during defibrillation causes mechanical forces (electrostriction), and as a result the heart deforms. This paper analyses the physical origin of the deformation, and how significant it is. Methods—We represent the heart as an anisotropic cylinder. This simple geometry allows us to obtain analytical solutions for the potential, current density, charge, stress, and strain. Results—Charge induced on the heart surface in the presence of the electric field results in forces that deform the heart. In addition, the anisotropy of cardiac tissue creates a charge density throughout the tissue volume, leading to body forces. These two forces cause the tissue to deform in a complicated manner, with the anisotropy suppressing radial displacements in favor of tangential ones. Quantitatively, the deformation of the tissue is small, although it may be significant when using some imaging techniques that require the measurement of small displacements. Conclusions—The anisotropy of cardiac tissue produces qualitatively new mechanical behavior during a strong, defibrillation-strength electric shock.

I.    BACKGROUND

The heart is normally controlled by an electrical signal that is organized and periodic. When fibrillation occurs this signal is no longer periodic, causing heart contraction to become rapid and disorganized. In order to stop fibrillation (that is, to defibrillate), a strong electric shock is applied to the heart that terminates this chaotic electrical behavior, resetting the heart to its normal rhythm [1]. This shock also affects the contraction of the cardiac muscle, influencing the way the heart moves. However, during defibrillation there is another force acting on the heart in addition to the contractile force normally generated by muscle fibers. The applied electric field produces mechanical forces by electrostriction: electrical forces acting on charge induced in the tissue.

Muscle has particularly interesting electrical properties because it is anisotropic: the electrical conductivity parallel to the muscle fibers is different than the conductivity perpendicular to them. This anisotropy not only has implications for the purely electrical behavior of the tissue, but also for its electrostrictive behavior [2]. Normally, electrostrictive effects are small. However, given the strong electric fields used during defibrillation and the anisotropic properties of cardiac muscle, we wondered if electrostriction might play any role during defibrillation, or in other applications in which large electric fields are applied.

The purpose of this paper is to determine the effect of electrostriction during defibrillation. Specifically, our goals are 1) to illustrate the basic physics underlying electrostriction in anisotropic cardiac tissue, 2) to calculate the contribution of electrostriction to how the heart deforms in response to an electric shock, and 3) to estimate the deformation caused by a shock during defibrillation.

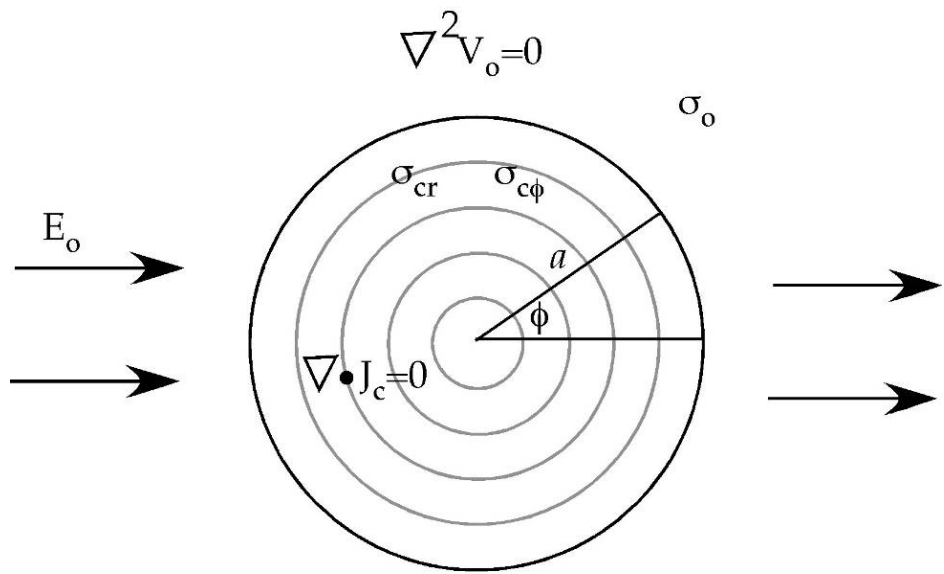

**Figure 1.** A schematic diagram of the heart in a uniform electric field. The gray circles indicate the myocardial fiber geometry.

II.    METHODS

Our general strategy is to analyze as simple a model of defibrillation as possible while still capturing the basic physics. In particular, we seek analytical solutions to the equations governing both the electrical and mechanical behavior. Our model could be made more realistic in many ways, at the price of making the solutions more complex.

a.    Electrical Model

We represent cardiac muscle as a macroscopic continuum, with the electrical properties of the tissue averaged over many cells. The orientation of the individual muscle fibers makes the macroscopic electrical conductivity anisotropic: the conductivity is greater along the fibers than perpendicular to them. Furthermore, we assume the electric potential is independent of time. Even though the capacitance of the cell membrane implies that the electrical properties of cardiac tissue are time dependent, the steady-state assumption should provide a good qualitative estimate of the potential at the end of a defibrillation shock. A typical shock lasts a few milliseconds, whereas the time constant of the membrane is about one millisecond.

The heart is approximated as a cylinder with radius $a$ immersed in an isotropic fluid of conductivity $\sigma_o$ (Fig. 1). A uniform electric field of strength $E_o$ is applied in the direction $\phi = 0$. We assume the myocardial fibers are in circular loops in the $\phi$ direction. The conductivity parallel to the fibers is $\sigma_{c\phi}$, and perpendicular to the fibers is $\sigma_{cr}$.

In the fluid, the potential $V_o$ obeys Laplace's equation, $\nabla^2 V_o = 0$. In the tissue, the anisotropy complicates the equation for the potential $V_c$. The fundamental equation is continuity of current, $\nabla \bullet J_c = 0$. The equation governing $V_c$ is therefore

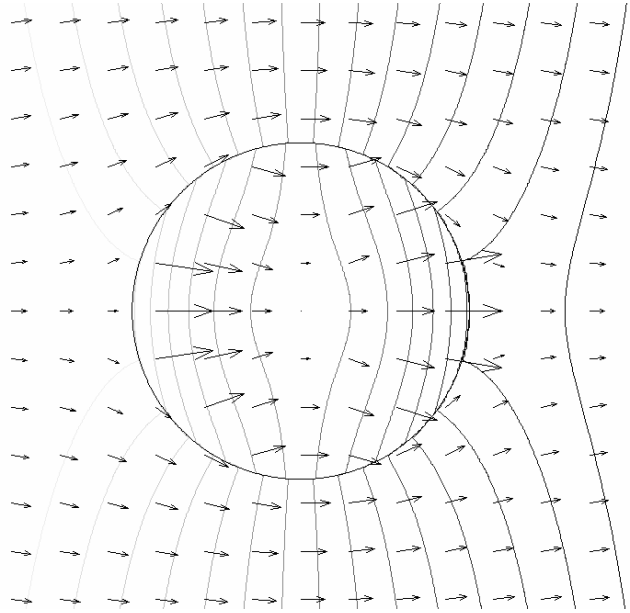

**Figure 2.** The potential (contour lines) and electric field (arrows) distributions in and around the heart during a shock, from Eqn. 4 and 5. $\lambda = 2$ and $\gamma = 2/3$. Light gray contours are positive, and dark gray contours are negative.

$$\sigma_{cr}\frac{1}{r}\frac{\partial}{\partial r}\left(r\frac{\partial V_c}{\partial r}\right)+\sigma_{c\phi}\frac{1}{r^2}\frac{\partial^2 V_c}{\partial \phi^2}=0 \ . \quad (1)$$

At the boundary $r = a$ the radial component of the current density **J** and the tangential component of the electric field **E** are continuous,

$$J_{or}=J_{cr} \quad \Rightarrow \quad \sigma_o\frac{\partial V_o}{\partial r}=\sigma_{cr}\frac{\partial V_c}{\partial r}, \quad (2)$$

$$E_{o\phi}=E_{c\phi} \quad \Rightarrow \quad V_o=V_c \ . \quad (3)$$

The solutions for the potentials are

$$V_o(r,\phi)=-E_o\, a\left(\left(\frac{a}{r}\right)\gamma+\left(\frac{r}{a}\right)\right)\cos\phi, \quad (4)$$

$$V_c(r,\phi)=-E_o\, a\left(\frac{r}{a}\right)^\lambda (\gamma+1)\cos\phi, \quad (5)$$

where $\lambda=\sqrt{\dfrac{\sigma_{c\phi}}{\sigma_{cr}}}$ and $\gamma=\left(\dfrac{\sigma_o-\lambda\sigma_{cr}}{\sigma_o+\lambda\sigma_{cr}}\right)$.

Figure 2 shows the potential and electric field.

The parameters $\lambda$ and $\gamma$ play important roles in the following analysis, so let us consider their values. The degree of anisotropy determines $\lambda$. Cardiac tissue has a value of about two[*]. Generally, the heart has a somewhat lower conductivity than the surrounding fluid, so $\gamma$ is positive between zero and one, with a typical value being about two-thirds (assuming $\sigma_o$ = 1 S/m).

---

[*] Cardiac tissue has different conductivities in the intracellular (inside the cells) (i) and extracellular (outside the cells) (e) space (it is a "bidomain"). In the monodomain model presented here, $\sigma_{c\phi}$ and $\sigma_{cr}$ should be thought of as the parallel combination of the intracellular and extracellular conductivities in each direction. The bidomain conductivities have been estimated from experiments, and are approximately [3] $\sigma_{i\phi}$=0.2 S/m, $\sigma_{e\phi}$=0.2 S/m, $\sigma_{ir}$=0.02 S/m, and $\sigma_{er}$=0.08 S/m. The anisotropy is then $\lambda=\sqrt{(\sigma_{i\phi}+\sigma_{e\phi})/(\sigma_{ir}+\sigma_{er})}=2$.

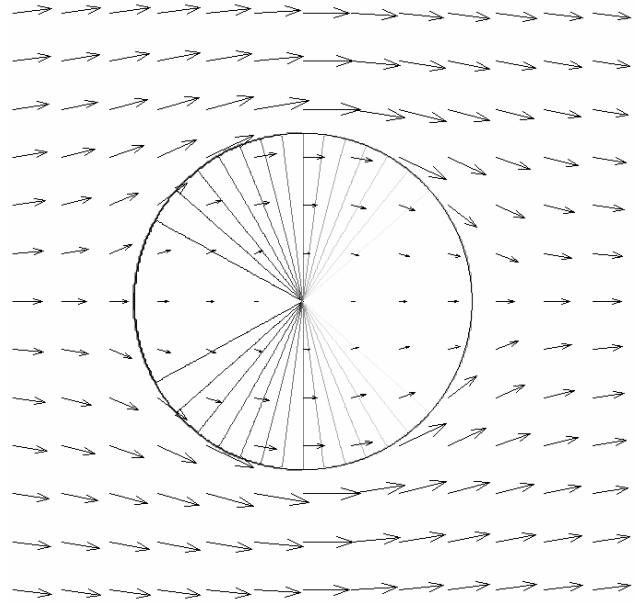

**Figure 3.** The charge density from Eqn. 7 (contour lines) and current density (arrows) in the heart during a shock. $\lambda = 2$ and $\gamma = 2/3$. Light gray contours are positive, and dark gray contours are negative.

A charge distribution builds up on the surface of the cylinder. In fact, it is this surface charge that ensures the boundary conditions at the interface are satisfied. The charge per unit area, *s*, is calculated from the electric field by $s = \varepsilon_o (E_{or} - E_{cr})$,

$$s = -\varepsilon_o E_o \cos\phi \left[\gamma(\lambda+1) + (\lambda-1)\right]. \quad (6)$$

Because the tissue is anisotropic, the continuity of current does not imply that the electric field has zero divergence. Gauss's Law relates the volume charge density $\rho$ to the electrical field **E**: $\rho = \varepsilon_o \nabla \bullet \mathbf{E}$. In our problem

$$\rho = \frac{\varepsilon_o E_o}{a}(\gamma+1)(\lambda^2-1)\left(\frac{r}{a}\right)^{\lambda-2} \cos\phi \ . \quad (7)$$

Figure 3 shows the charge density and the current density.

We can integrate the surface charge over the right half the cylinder surface (integrating *s* over the entire surface would cause it to vanish). For a section of the cylinder of length *L*, the surface charge is $Q_{surface} = -\varepsilon_o E_o 2La[\gamma(\lambda+1) + (\lambda-1)]$. Integrating $\rho$ over the right half of the cylinder volume gives the charge distributed throughout the tissue volume, $Q_{volume} = \varepsilon_o E_o 2La(\gamma+1)(\lambda^2-1)/\lambda$.

For typical values of $\lambda$ and $\gamma$, $Q_{surface}$ and $Q_{volume}$ have similar magnitudes, but opposite signs. Of course, $Q_{volume}$ goes to zero if $\lambda$ equals one (signifying isotropic tissue).

b. Mechanical Model

In general, the charge distribution in the tissue interacts with the electric field to produce a force. In particular, the volume

$$F_{re} = \frac{\varepsilon_o E_o^2}{2a}\left(\frac{r}{a}\right)^{2\lambda-3}(\gamma+1)^2(\lambda^2-1)\lambda\,(1+\cos 2\phi)\;, \tag{8}$$

$$F_{\phi e} = -\frac{\varepsilon_o E_o^2}{2a}\left(\frac{r}{a}\right)^{2\lambda-3}(\gamma+1)^2(\lambda^2-1)\sin 2\phi\;. \tag{9}$$

charge density results in a body force (force per unit volume) $\mathbf{F}=\rho\mathbf{E}$. Figure 4 shows the body force. The body force, in addition to the force on the surface charge, causes the tissue to deform.

Because the tissue is mostly water, we assume that it is incompressible $(\nabla\bullet\mathbf{u}=0)$, so the displacement $\mathbf{u}$ can be given in terms of a stream function [4] $\psi$

$$u_r = -\frac{1}{r}\frac{\partial\psi}{\partial\phi}\;, \tag{10}$$

$$u_\phi = \frac{\partial\psi}{\partial r}\;. \tag{11}$$

Love[4] gives the strain tensor $\varepsilon_{ij}$ in terms of the displacement in polar coordinates

$$\varepsilon_{rr} = \frac{\partial u_r}{\partial r}\;, \tag{12}$$

$$\varepsilon_{\phi\phi} = \frac{u_r}{r} + \frac{1}{r}\frac{\partial u_\phi}{\partial\phi}\;, \tag{13}$$

$$\varepsilon_{r\phi} = \frac{1}{2}\left(\frac{1}{r}\frac{\partial u_r}{\partial\phi}+\frac{\partial u_\phi}{\partial r}-\frac{u_\phi}{r}\right). \tag{14}$$

The stress tensor $\tau_{ij}$ and the strain tensor are related by

$$\tau_{ij} = -p\delta_{ij} + 2\mu\varepsilon_{ij} + T_{ij}\;. \tag{15}$$

The first term represents the fluid hydrostatic pressure, $p$ (where $\delta_{ij}$ is the Kronecker delta). The second term is a Hookean relationship between shear stress and shear strain, with $\mu$ being the shear modulus [5]. The third term is the electrostatic part of the Maxwell stress tensor [6], $T_{ij}=\varepsilon_o\left(E_i E_j - \delta_{ij}E^2/2\right)$. With this stress-strain relationship, the equations for mechanical equilibrium (Navier's equation) in polar coordinates are [4]

$$-\frac{\partial p}{\partial r}+2\mu\left(\frac{\partial\varepsilon_{rr}}{\partial r}+\frac{1}{r}\frac{\partial\varepsilon_{r\phi}}{\partial\phi}+\frac{\varepsilon_{rr}-\varepsilon_{\phi\phi}}{r}\right)+F_r=0, \tag{16}$$

$$-\frac{1}{r}\frac{\partial p}{\partial\phi}+2\mu\left(\frac{\partial\varepsilon_{r\phi}}{\partial r}+\frac{1}{r}\frac{\partial\varepsilon_{\phi\phi}}{\partial\phi}+\frac{2\varepsilon_{r\phi}}{r}\right)+F_\phi=0. \tag{17}$$

At the free boundary ($r = a$) $\tau_{rr}$ and $\tau_{r\phi}$ are continuous [4].

III. RESULTS

The pressure $p$ and the stream function $\psi$ are

$$p_i = \frac{\varepsilon_o E_o^{\,2}(\gamma+1)^2}{4}\left[\lambda(\lambda+1)\left(\frac{r}{a}\right)^{2\lambda-2}-\lambda-1+\frac{4\gamma}{(\gamma+1)^2}\right.$$
$$\left.+\cos 2\phi\left(\frac{(\lambda^2-1)(\lambda^2-\lambda-1)}{\lambda(\lambda-2)}\left(\frac{r}{a}\right)^{2\lambda-2}-\left(\frac{3(\lambda-1)(\lambda^2-\lambda-1)}{\lambda(\lambda-2)}+\frac{4\gamma^2}{(\gamma+1)^2}\right)\left(\frac{r}{a}\right)^2\right)\right], \tag{18}$$

$$p_o = 0. \tag{19}$$

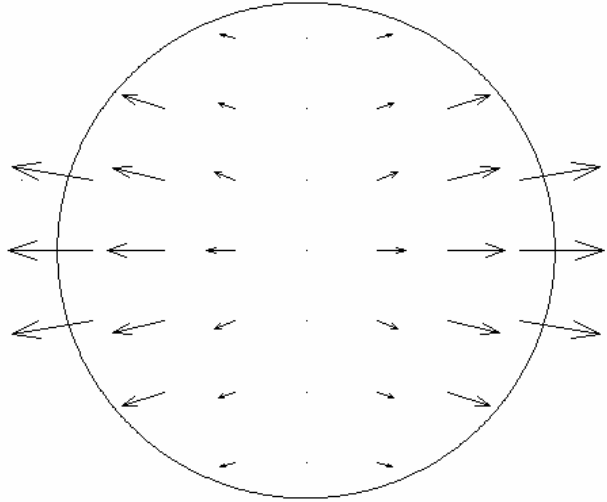

**Figure 4.** The distribution of the body force during a shock, from Equations 8 and 9. The force vector is indicated by the arrows, and is produced by charge interacting with the electric field throughout the tissue. $\lambda = 2$ and $\gamma = 2/3$.

$$\psi_i = -\frac{\varepsilon_o E_o^2 a^2}{16\mu}(\gamma+1)^2 \sin 2\phi$$

$$\left[\frac{1}{\lambda(\lambda-2)}\left(\frac{r}{a}\right)^{2\lambda} - \left(\frac{(\lambda-1)(\lambda^2-\lambda-1)}{\lambda(\lambda-2)} + \frac{4\gamma^2}{3(\gamma+1)^2}\right)\left(\frac{r}{a}\right)^4 + \left(\frac{\lambda^2-\lambda-1}{\lambda} + \frac{2(\gamma^2+1)}{(\gamma+1)^2}\right)\left(\frac{r}{a}\right)^2\right],$$

(20)

$$\psi_o = 0. \tag{21}$$

From the stream function we can determine the radial and angular displacement

$$u_r = \frac{\varepsilon_o E_o^2 a}{8\mu}(\gamma+1)^2 \cos 2\phi$$

$$\times \left[\frac{1}{\lambda(\lambda-2)}\left(\frac{r}{a}\right)^{2\lambda-1} - \left(\frac{(\lambda-1)(\lambda^2-\lambda-1)}{\lambda(\lambda-2)} + \frac{4\gamma^2}{3(\gamma+1)^2}\right)\left(\frac{r}{a}\right)^3 + \left(\frac{\lambda^2-\lambda-1}{\lambda} + \frac{2(\gamma^2+1)}{(\gamma+1)^2}\right)\left(\frac{r}{a}\right)\right],$$

(22)

$$u_\phi = -\frac{\varepsilon_o E_o^2 a}{8\mu}(\gamma+1)^2 \sin 2\phi$$

$$\left[\frac{1}{(\lambda-2)}\left(\frac{r}{a}\right)^{2\lambda-1} - \left(\frac{2(\lambda-1)(\lambda^2-\lambda-1)}{\lambda(\lambda-2)} + \frac{8\gamma^2}{3(\gamma+1)^2}\right)\left(\frac{r}{a}\right)^3 + \left(\frac{\lambda^2-\lambda-1}{\lambda} + \frac{2(\gamma^2+1)}{(\gamma+1)^2}\right)\left(\frac{r}{a}\right)\right].$$

(23)

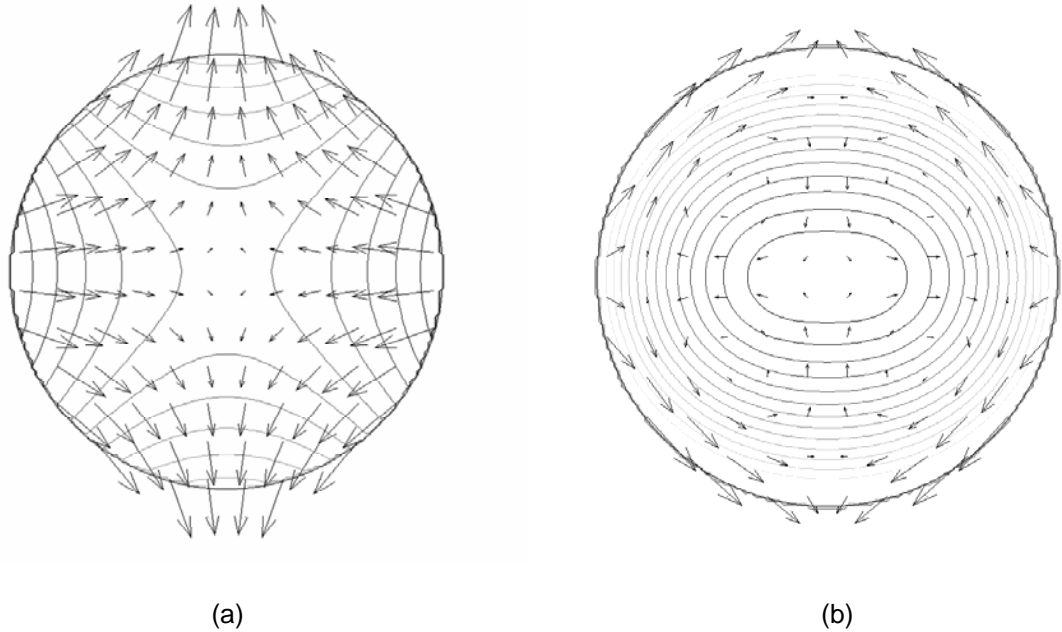

(a) (b)

**Figure 5.** The pressure (contour lines) and displacement (arrows) in the heart during a shock, based on Eqn. 18 – 23: a) Isotropic tissue, $\lambda = 1$ and $\gamma = 2/3$; b) typical heart tissue, $\lambda = 2$ and $\gamma = 2/3$. Light gray contours are positive, and dark gray contours are negative.

The pressure and displacement are shown in Figure 5(a) for isotropic tissue, and Figure 5(b) for normal, anisotropic cardiac tissue.

IV. DISCUSSION

Two features of the deformation (Equations 22 and 23) are of interest: its qualitative distribution, and its quantitative magnitude. Qualitatively, the surface charge compresses the cylinder along the direction of the electric field. However, the body force arising from the volume charge has the opposite effect, elongating the cylinder along the direction of the field (Fig. 4). The net result of these two opposing actions is to reduce the radial deformation in the tissue, Figure 5(b), compared to the isotropic case, Figure 5(a), while maintaining the tangential deformation. Moreover, in the inner core of the cylinder, the radial displacement changes sign from inward to outward because of the anisotropy--compare Figures 5(a) and 5(b). Thus, the deformation in anisotropic tissue is rather complex, and qualitatively different than in isotropic tissue.

The equations for the pressure, stream function, and displacement appear to diverge for $\lambda = 2$, because of the factor of $\lambda - 2$ in the denominator. Actually, the functions are well behaved. For instance, in the limit as $\lambda$ goes to 2, the radial displacement becomes

$$u_r = \frac{\varepsilon_o E_o^2 a}{8\mu}(\gamma+1)^2 \cos 2\phi \left[\left(-2 + \frac{4\gamma^2}{3(\gamma+1)^2}\right)\left(\frac{r}{a}\right)^3 + \left(\frac{1}{2} + \frac{2(\gamma^2+1)}{(\gamma+1)^2}\right)\left(\frac{r}{a}\right)\right]. \qquad (24)$$

The charge density (Eqn. 7) diverges at $r=0$ if $\lambda < 2$. This behavior arises because of the small radius of curvature of the fibers at the origin. In a more realistic model, in which the fibers have a finite curvature, this effect would probably not occur.

Quantitatively, electrostrictive effects are small. Equation 18 gives the pressure as $\varepsilon_o E_o^2$ times a dimensionless factor that depends on the tissue parameters and position. Niemann et al. [7] estimate that electric fields during defibrillation can be on the order of 3000 V/m, implying a pressure of $80 \times 10^{-6}$ Pa. Equations 22 and 23 indicate that the displacement is $\varepsilon_o E_o^2 a / \mu$ times a dimensionless factor. If the heart has a radius of 0.04 m and the shear modulus is 5000 Pa, the displacement is $0.64 \times 10^{-9}$ m. Clearly, electrostrictive effects are much smaller than the usual muscle contraction associated with the cardiac action potential. They are also too small to explain unusual mechanical deformations observed by Sylvester *et al.* [8] during defibrillation. However, some imaging methods record very small displacements and pressures, such as Magneto-acoustic imaging [9-12], Magneto-acoustic Tomography with Magnetic Induction, [13,14] Magnetic Resonance Elastography, [15-17] and Lorentz Force imaging [18-21]. In these imaging techniques, electrostrictive effects could contribute to the measured signal.

Our study has some significant limitations. The tissue is represented as a macroscopic continuum [22], so we do not account for the detailed behavior at the level of a single cell. Our steady-state model does not capture the time dependence of the process. The heart shape is only approximately cylindrical, and the fiber geometry in the heart is more complex than assumed in our calculation [23]. We use a simple representation of a circular fiber geometry in order to obtain an analytical solution. We ignore any differences of tissue properties between different parts of the heart. The heart has an inner cavity filled with blood that is not present in our calculation, but could be incorporated into an improved version of the model. The divergence of the charge density at r = 0 would disappear in that case. We assume the mechanical properties of the tissue are isotropic [5], although the electrical conductivity is anisotropic. The tissue is represented as a single conducting medium, although cardiac tissue is better represented by a bidomain, in which the intracellular and extracellular spaces are accounted for individually. Nevertheless, our simple geometry has the virtue of allowing an analytical solution for the pressure and displacement. Our analytic equations show how the solution depends on the model parameters, and provide a known solution that can be used to test more realistic numerical calculations. Another advantage of an analytical solution is that it illustrates the basic physics underlying electrostriction in anisotropic tissue.

## V. CONCLUSIONS

In this paper we have accomplished three goals. 1) We have illustrated the basic physics of electrostriction in the heart. Anisotropic tissue such as cardiac muscle has a unique electrostrictive behavior. A charge density is present throughout the tissue, resulting in body forces that contribute to the tissue deformation. 2) We have calculated how the heart deforms in response to a shock. Figure 5 shows that the heart shortens along the direction of the electric field, and lengthens in the direction perpendicular to it. The anisotropy qualitatively changes the distribution of displacements, compared to the isotropic case. 3) We have determined that the effect of electrostriction is small. It is therefore likely it does not play a significant role in the mechanical deformations of the heart during defibrillation. However, electrostriction effects are large enough that they could improve the accuracy of existing imaging modalities that measure small displacements.

## ACKNOWLEDGEMENTS

This research was supported by the SMaRT program at Oakland University, a Research Experience for Undergraduates (REU) and was funded by NSF grant DMR-055 2779.